\documentclass[twocolumn]{aastex631}
\usepackage{amsmath}

\usepackage{ulem}

\newcommand{\msun}{${\rm M_{\sun}}$}
\def\ltsima{$\; \buildrel < \over \sim \;$}
\def\simlt{\lower.5ex\hbox{\ltsima}}
\def\gtsima{$\; \buildrel > \over \sim \;$}
\def\simgt{\lower.5ex\hbox{\gtsima}}
\def\km{{\rm\,km}}
\def\kms{{\rm\,km\,s^{-1}}}
\def\km2s2{{\rm\,km^2\,s^{-2}}}
\def\pc{{\rm\,pc}}
\def\kpc{{\rm\,kpc}}

\def\msun{{\rm\,M_\odot}}


\def\deg{^\circ}

\def\Gyr{{\rm\,Gyr}}

\def\masyr{{\rm\,mas \, yr^{-1}}}
\def\sLz{{\sigma_{L_z}}}
\def\sv{{\sigma_{v_\mathrm{los}}}}
\def\vlos{{{v_\mathrm{los}}}}
\def\svtan{{\sigma_{v_\mathrm{Tan}}}}
\def\svtans{{\sigma^{s}_{v_\mathrm{Tan}}}}
\def\vtan{v_\mathrm{Tan}}

\def\ltsima{$\; \buildrel < \over \sim \;$}
\def\gtsima{$\; \buildrel > \over \sim \;$}

\defcitealias{MalhanCocoonDetection2019}{M19a}
\defcitealias{Malhan2019_GD1_Kshir}{M19b}
\defcitealias{Bonaca2019Jhelum}{B19}
\defcitealias{Carlberg2018Density_Structure}{C18}
\defcitealias{Malhan2018PotentialGD1}{MI19}
\def\Malhan_ppr{{Malhan et al. [submitted]}}

\def\Gaia{{\it Gaia}}

\def\KF#1{\noindent{{\color{blue} [KF: #1]}}}
\def\KF2#1{\noindent{{\color{cyan} [KF2: #1]}}}

\submitjournal{ApJL}

\shorttitle{Probing dark matter using accreted globular cluster streams}
\shortauthors{Malhan et al.}
\graphicspath{{./}{figures/}}

\begin{document}

\title{New constraints on the dark matter density profiles of dwarf galaxies from proper motions of globular cluster streams}

\correspondingauthor{Khyati Malhan}
\email{kmalhan07@gmail.com}

\author[0000-0002-8318-433X]{Khyati Malhan}
\affiliation{Humboldt Fellow and IAU Gruber Fellow}
\affiliation{The Oskar Klein Centre, Department of Physics, Stockholm University, AlbaNova, SE-10691 Stockholm, Sweden}
\affiliation{Max-Planck-Institut f\"ur Astronomie, K\"onigstuhl 17, D-69117, Heidelberg, Germany}

\author[0000-0002-6257-2341]{Monica Valluri}
\affiliation{Department of Astronomy, University of Michigan, Ann Arbor, MI, 48109, USA}

\author[0000-0001-9490-020X]{Katherine Freese}
\affiliation{The Oskar Klein Centre, Department of Physics, Stockholm University, AlbaNova, SE-10691 Stockholm, Sweden}
\affiliation{Theory Group, Department of Physics, The University of Texas at Austin, 2515 Speedway, C1600, Austin, TX 78712-0264, USA}

\author[0000-0002-3292-9709]{Rodrigo A. Ibata}
\affiliation{Universit\'e de Strasbourg, CNRS, Observatoire astronomique de Strasbourg, UMR 7550, F-67000 Strasbourg, France}

%
\begin{abstract}

The central density profiles in dwarf galaxy halos depend strongly on the nature of dark matter. Recently, in Malhan et al. (2021), we employed N-body simulations to show that the cuspy cold dark matter (CDM) subhalos predicted by cosmological simulations can be differentiated from cored subhalos using the properties of accreted globular cluster (GC) streams since these GCs experience tidal stripping within their parent halos prior to accretion onto the Milky Way.  We previously found that clusters that are accreted within cuspy subhalos produce streams with larger physical widths and higher dispersions in line-of-sight velocity and angular momentum than streams that are accreted within cored subhalos. Here, we use the same suite of simulations to demonstrate that the dispersion in the tangential velocities of streams ($\svtan$) is also sensitive to the central DM density profiles of their parent dwarfs and GCs that were accreted from: cuspy subhalos produce streams with larger $\svtan$ than those accreted inside cored subhalos. Using {\it Gaia} EDR3 observations of multiple GC streams we compare their ${\svtan}$ values with simulations.  The measured $\svtan$ values are consistent with both an ``in situ'' origin and with accretion inside cored subhalos of $M\sim10^{8-9}M_{\odot}$ (or very low-mass cuspy subhalos of mass $\sim10^8M_{\odot}$). Despite the large current uncertainties in $\svtan$, we find a low probability that any of the progenitor GCs were accreted from cuspy subhalos of $M\gtrsim 10^9 M_{\odot}$. The uncertainties on {\it Gaia} tangential velocity measurements are expected to decrease in future and will allow for stronger constraints on subhalo DM density profiles.

\end{abstract}

\section{Introduction}\label{sec:Introduction}

The true nature of dark matter (DM) is currently unknown (cf.  \citealt{Bertone2005_Particle_DM_review}) and our understanding about this mysterious particle is based primarily on theoretical predictions from cosmological simulations and observations of large scale structure (cf. \citealt{Salucci_2019}). While particle physicists have been working for decades to set limits on the mass of the putative DM particle, much is still unknown. For instance, the widely accepted cold dark matter (CDM) theory hypothesizes that the DM particle is non-relativistic (``cold''), collisionless and weakly interacting \citep{White_CDM_candidate1978, Blumenthal1984}. The CDM framework predicts that galaxy halos (irrespective of their sizes) should possess {\it cuspy} DM distributions, with very steeply rising inner density profiles of the form $\rho_{\rm DM}\propto r^{-1}$ \citep{Dubinski1991, Navarro_1997}. Alternative theories which differ from CDM in terms of the behaviour of their elementary particles (e.g. ultra-light DM, a.k.a. fuzzy DM, \citealt{Hui2017_ULDM_FDM}), interaction strength (e.g. self-interacting DM, \citealt{Spergel2000SIDM, Elbert2015}), etc. Interestingly, many of these alternative DM theories instead predict {\it cored} DM distributions on galactic/sub-galactic scales, where central densities are approximately constant. Therefore, measurements of the central DM densities in dwarf galaxies provide a possible avenue to constrain the fundamental properties of DM. It has also been previously suggested that the widths of dwarf galaxy streams can also be a probe of their parent galaxy's DM density  profiles \citep{Errani_2015}.

Recently, in \cite{Malhan_DM_2021}, we presented a new method of probing the central DM densities in dwarf galaxies using  globular cluster (GC) stellar streams.
Stellar streams are produced from the tidal stripping of a progenitor  (e.g. a GC) as it orbits in the potential of the host galaxy. In the Milky Way (MW), 
nearly a 100 streams have been detected to date
\citep{Mateu_22}. 
Among this set, some of the progenitor GCs  of these streams are suspected to have been accreted; i.e., these GC streams originally evolved within their parent dwarf galaxies and only later merged with the MW (e.g., \citealt{Malhan_cocoon_2019, Malhan_Kshir_2019, Bonaca_2021, Malhan_2022}). Motivated by this scenario, we asked in \cite{Malhan_DM_2021}: {\it can the present day physical properties of accreted GC streams inform us about the DM density profiles inside their parent dwarf galaxies?} To explore this question, we ran several N-body simulations and showed that GCs that accrete within cuspy CDM subhalos produce streams that are substantially wider (physically) and dynamically hotter than those streams that accrete inside cored subhalos. This difference occurs due to the difference in the dynamical evolution of GCs inside two different potential models -- cuspy and cored -- with the former causing larger tidal stripping of the GC (inside the parent subhalo) than the latter. This implies that the physical properties of accreted GC streams provides a means to probe the DM density profiles inside their parent dwarfs.

In \cite{Malhan_DM_2021}, the physical properties of the streams were quantified in terms of their a) transverse physical widths ($w$), b) dispersion in the line-of-sight (los) velocities ($\sv$), and c) dispersion in the z-component of angular momenta ($\sLz$). We found that these parameters differ in in cuspy and cored halos (see Figure~7 of \citealt{Malhan_DM_2021}). In particular, the parameters $\sv$ and $\sLz$ depend on the spectroscopic los velocities, that we lack for a majority of stream stars. However, with ESA/\Gaia\ mission \citep{Prusti_2016}, we now possess excellent proper motions and parallaxes for millions of halo stars, and this data can be used to measure the tangential velocities ($\vtan$) of stream stars. Our aim in this work is to show that the intrinsic dispersion in the tangential velocities of stream stars ($\svtan$) can be used as an alternative parameter to differentiate between the cusp/core scenario (at least for halos of mass$\simgt10^9\msun$), and this provides a new means to probe the central DM density profiles inside the dwarf galaxies. 

This article is arranged as follows. Section~\ref{sec:Tan_vel_of_simulations} details the computation of $\svtan$ for the simulated stream models produced in cuspy vs. cored halos. Section~\ref{sec:Tan_vel_of_observed_streams} describes the procedure to measure $\svtan$ of the observed streams of the MW using \Gaia\ EDR3 dataset \citep{GaiaEDR3_Lindegren_2020}. Finally, in Section~\ref{sec:Conclusion}, we compare $\svtan$ of the observations and the simulations and provide our conclusions.

\section{Tangential velocity dispersions of the N-body stream models}\label{sec:Tan_vel_of_simulations}


%
\subsection{N-body simulations of accreted globular cluster streams}\label{subsec:simulation}

The N-body stream simulations in \cite{Malhan_DM_2021} were of two types: those that were produced by {\it in situ} GCs and those that were produced {\it accreted} GCs. 

The in situ GC streams arise from GCs that formed inside the MW, and whose evolution has been primarily determined by the MW potential. In \cite{Malhan_DM_2021}, we simulated $n=5$ in situ GC streams. The progenitor GC were modeled by King profiles \citep{King1962} with masses ranging from $M_{\rm GC}=[3-10]\times10^{4}\msun$, central potential ranging from $W=1.5-3$ and tidal radius from $r_t = 0.05-0.2\kpc$. This mass range was motivated by previous studies on clusters and streams of the Milky Way (e.g., \citealt{Baumgardt_2016, Thomas_Pal52016}). The star particles had individual masses of $5\msun$ and softenings of $2\pc$. To evolve these N-body GC models in a host Galactic potential (that mimics the MW), we used model \#1 of \cite{Dehnen1998Massmodel}. This is a static, axisymmetric potential comprising of a thin disk, a thick disk, interstellar medium, bulge and DM halo. The simulations were evolved for $T= 8\Gyr$ using the collisionless \texttt{GyrfalcON} integrator \citep{Dehnen_NEMO_2002} from the \texttt{NEMO} package \citep{Teuben_1995}. 

To produce accreted GC streams, we tried a total of $4$ parent subhalos that were constructed using the Dehnen model \citep{Dehnen1993}. The Dehnen model is expressed as 
\begin{equation}\label{eq:density_dehnen}
\rho (r) =\frac{(3-\gamma)M_0}{4\pi r^3_0} \Big( \frac{r}{r_0} \Big)^{-\gamma} \Big(1+ \frac{r}{r_0} \Big)^{\gamma-4},
\end{equation}
where $M_0, r_0,- \gamma$ are the mass, scale radius, and the logarithmic slope of the inner density profile of the subhalo, respectively. Two of the subhalos possess cuspy (NFW-like) profile and two possess cored density profiles. These subhalos are described as 1) SCu (small/cuspy) model: $\{M_0,r_0,\gamma\}$=$\{10^8\msun, 0.75\kpc,1\}$; 2) SCo (small/cored) model: $\{M_0,r_0,\gamma\}$=$\{10^8\msun, 0.75\kpc,0\}$; 3) LCu (large/cuspy) model: $\{M_0,r_0,\gamma\}$=$\{10^9\msun, 1.60\kpc,1\}$ and 4) LCo (large/cored) model: $\{M_0,r_0,\gamma\}$ = $\{10^9\msun, 1.60\kpc,0\}$. This mass range was adopted because it is similar to the masses of some of the dwarf galaxies that host GCs (e.g., \citealt{Forbes_2018}), and also similar to the mass of the (hypothesized) parent dwarf galaxy of the ``GD-1'' stream \citep{Malhan_Kshir_2019,Malhan_cocoon_2019}. The mass and softening parameters of the DM particles were $750\msun$ and $20\pc$, respectively\footnote{This choice of resolution, for both the subhalos and the GCs, was based on several numerical tests that we undertook in \cite{Malhan_DM_2021}.}. Each subhalo model was populated with one GC model, and this GC was placed at an off-centre location and was launched on an orbit inside the subhalo. At the same time, the subhalo was launched on an orbit inside the host Galactic potential. The integration time of every simulation was $T= 8\Gyr$ and the GC spends $\sim3-4\Gyr$ inside the parent subhalo before escaping into the host (see \citealt{Malhan_DM_2021}).  

We ran over $100$ N-body simulations of accreted GC streams, including many different orbital configurations of GCs inside the subhalo (see Table~1 of \citealt{Malhan_DM_2021}). The majority of orbits of the subhalos (hosting the GC) within the MW, were circular (with galactocentric radius  $\sim60\kpc$), and only a few were eccentric. Furthermore, while most of the simulations employed subhalos that lacked an extended population of stars, we did experiment with a few cases that included a stellar population (as expected from dwarf galaxies). However, we found that in both the cases (with and without the stellar population), the final morphologies of the accreted GC streams were the same. 

All of the GC stream models were transformed from the galactocentric Cartesian coordinates to the heliocentric equatorial coordinates from which we measure $\svtan$ of the simulated streams. This transformation provided for every star particle its position ($\alpha,\delta$), heliocentric distances ($d_{\odot}$) and proper motions ($\mu^{*}_{\alpha} \equiv \mu_{\alpha} {\rm cos} \delta,\mu_{\delta}$)\footnote{Naturally we also obtained $v_{\rm vlos}$ for each star, but that is not used in this study.}. Here, we use all of these quantities to measure $\svtan$ of streams. Note that these are the same quantities that are provided by the \Gaia\ dataset, except for $d_{\odot}$ (as \Gaia\ provides only parallaxes of stars). In Section~\ref{sec:Tan_vel_of_observed_streams} we discuss how  we use \Gaia\ parallaxes to estimate the distances to stream stars. 

\begin{figure*}
\begin{center}
\vspace{-0.3cm}
\includegraphics[width=0.7\hsize]{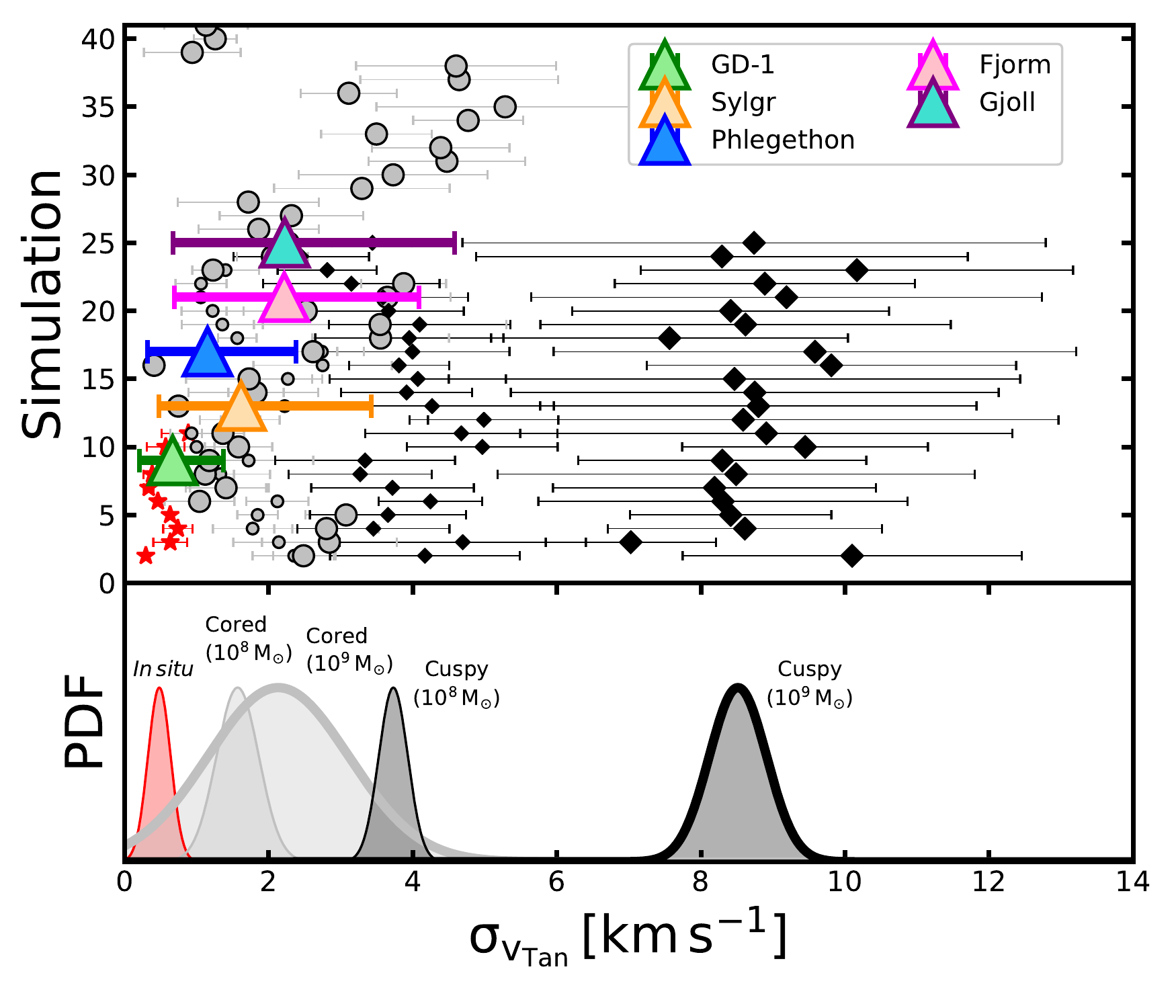}
\end{center}
\vspace{-0.8cm}
\caption{Using $\svtan$ of GC streams to probe the DM density profiles inside their parent subhalos (or parent dwarf galaxies). {\it Upper panel}: each red/black/gray point represents the tangential velocity dispersion ($\svtan$) of a particular simulated stream, and the corresponding error bar reflects the dispersion in the $\svtan$ measurement along that stream. The $Y$ axis denotes different simulations. The red points correspond to the {\it in situ} GC stream models and the black/gray points correspond to streams that {\it accreted} inside cuspy/cored subhalos (where small/large markers correspond to cases where subhalos had mass of $M=10^8\msun/10^9\msun$). The colored triangles are $\svtan$ values we measure for 5 Milky Way streams, using \Gaia\ EDR3 data. {\it Lower panel}:  red/black/gray Gaussians correspond respectively to the distribution of simulated $\svtan$ values from the in situ/cuspy/cored scenarios (including the scatter in the $\svtan$ measurements). Gaussian with thin/thick borders correspond to cases where subhalos had mass of $M=10^8\msun/10^9\msun$). In summary, in situ GC streams (red stars) possess extremely low values of $\svtan$, GC streams accreted inside cuspy CDM subhalos (black diamonds) possess very large values of $\svtan$, while streams accreted inside cored subhalos (gray circles) lie in between.}
\label{fig:Fig_Summary}
\end{figure*}
\subsection{Computing tangential velocity dispersion ($\svtan$) of simulated streams}\label{subsec:simulation_svtan}

To compute the dispersion in the tangential velocity of a given stream ($\svtan$), we first compute tangential velocities of the individual member stars ($\vtan$). Tangential velocity is defined as $\vtan=k \times d_{\odot} \times \mu$; where $k=4.7405\kms \kpc^{-1}(\masyr)^{-1}$, $\mu=\sqrt{\mu^{*\,2}_{\alpha} + \mu^2_{\delta}}$. Instead of computing $\svtan$, one maybe tempted to directly compute the dispersion in the proper motions; since it is the proper motion of stars that are provided by the \Gaia\ dataset. However, proper motions are distance dependent, therefore we use the dispersion in tangential velocities which is independent of distance. 

To compute $\svtan$ of simulated streams, we follow a similar approach as that used in \cite{Malhan_DM_2021} to measure other dynamical quantities. For a given stream, we first transform the positions of its member stars from the equatorial ($\alpha,\delta$) coordinate system to the ($\phi_1,\phi_2$) coordinate system, where $\phi_1$ is the angle that is aligned with the stream and $\phi_2$ is the angle perpendicular to the stream. Next, we consider small segments along $\phi_1$ of length $30\deg$ and compute $\svtans_{,i}$ independently for each $i^{\rm th}$ segment (the reason for undertaking this ``segment-wise'' calculation is described below). To compute $\svtans_{,i}$ in a given segment we first fit $\vtan$ of the star particles using a smooth function of the form
\begin{equation}\label{eq:vtan_fit}
\vtan (\phi_1) = a_1 + b_1 \phi_1 + c_1 \phi^2_1\, ,
\end{equation}
where $a_1,b_1,c_1$ are the fitting parameters to obtain the systemic value of $\vtan (\phi_1)$. After this, we subtract the fitted $\vtan$ function from the $\vtan$ of star particles to obtain the residual distribution. The standard deviation of this distribution provides the tangential velocity dispersion for the $i^{\rm th}$ segment of the stream (i.e., $\svtans_{,i}$). This procedure is iterated over all the segments in a given stream. Finally, the median and the standard deviation of the $\sigma^s_{{\rm vtan},i}$ distribution provides the $\svtan$ measurement for the entire stream and the dispersion on this measurement, respectively.  We use this procedure to compute $\svtan$ for all the N-body stream models. The reason we compute $\svtan$ independently for each segment of a stream is that many accreted GC streams are long and highly complex in structure (see Figures~1, 5, 6 of \citealt{Malhan_DM_2021}). Therefore, it is difficult to approximate the entire stream with a single function. Nonetheless, our procedure to obtain $\svtan$ also provides the dispersion on the $\svtan$ measurements. 

Figure~\ref{fig:Fig_Summary} (upper panel) shows the $\svtan$ measurements for all the N-body stream models as red stars (in-situ), grey circles (cored) and black diamonds (cuspy); the dispersions on their $\svtan$ are shown with error-bars. A visual inspection of this figure already indicates that streams produced in different scenarios (i.e. in situ, cuspy and cored) possess quite different values of $\svtan$. For a given in situ/cored/cuspy scenario, we quantify the variance in $\svtan$ distribution (denoted as $\langle \svtan  \rangle$) by modeling the $\svtan$ measurements with a Gaussian function of mean $\langle{x}\rangle$ and intrinsic dispersion $\sigma_x$. To this end, we use the MCMC sampler \texttt{emcee} \citep{emcee_2013} and define the log-likelihood function for every stream seqment $i$ as: 
\begin{equation}\label{eq:likelihood}
\begin{aligned}
    \ln \mathcal{L} &= \sum_i^n \left[ -\ln (\sqrt{2\pi}\sigma_i) -0.5\dfrac{(x_i - \langle{x}\rangle)^2}{\sigma_i^2}\right]\,,\\
     {\rm with}\,\, \sigma^2_i &=\sigma^2_{x} + \delta^2_i.
\end{aligned}
\end{equation}

Here, $x_i = \svtan$ of the stream segment $i$ and $\delta_i$ is the dispersion on $\svtan$ of that segment of the stream.

For the in situ GC streams, we find $\langle \svtan  \rangle= 0.5\pm0.1\kms$, implying that these streams are dynamically very cold. Furthermore, the cuspy SCu (LCu) subhalo with mass $M_0= 10^8(10^9)\msun$ produced GC streams with value $ \langle\svtan \rangle = 3.7\pm0.2 \, (8.5\pm0.4)\kms$. This implies that these streams are dynamically very hot. For the cored SCo (LCo) subhalo we infer the value of $ \langle\svtan \rangle = 1.6\pm0.3 \, (2.1\pm 1.0)\kms$. This difference in the $ \langle\svtan \rangle$ measurement of streams produced under cuspy/cored subhalo implies that present day $\svtan$ of streams are sensitive to the gravitational potential of their parent subhalos (i.e., $\svtan \propto M_0/r_0$). Some degeneracy between subhalo mass and the presence of a cusp is apparent in Figure~\ref{fig:Fig_Summary}. This issue is discussed further in Section~\ref{sec:Conclusion}.

\begin{figure*}
\begin{center}
\vspace{-0.1cm}
\vbox{
\includegraphics[width=0.32\hsize]{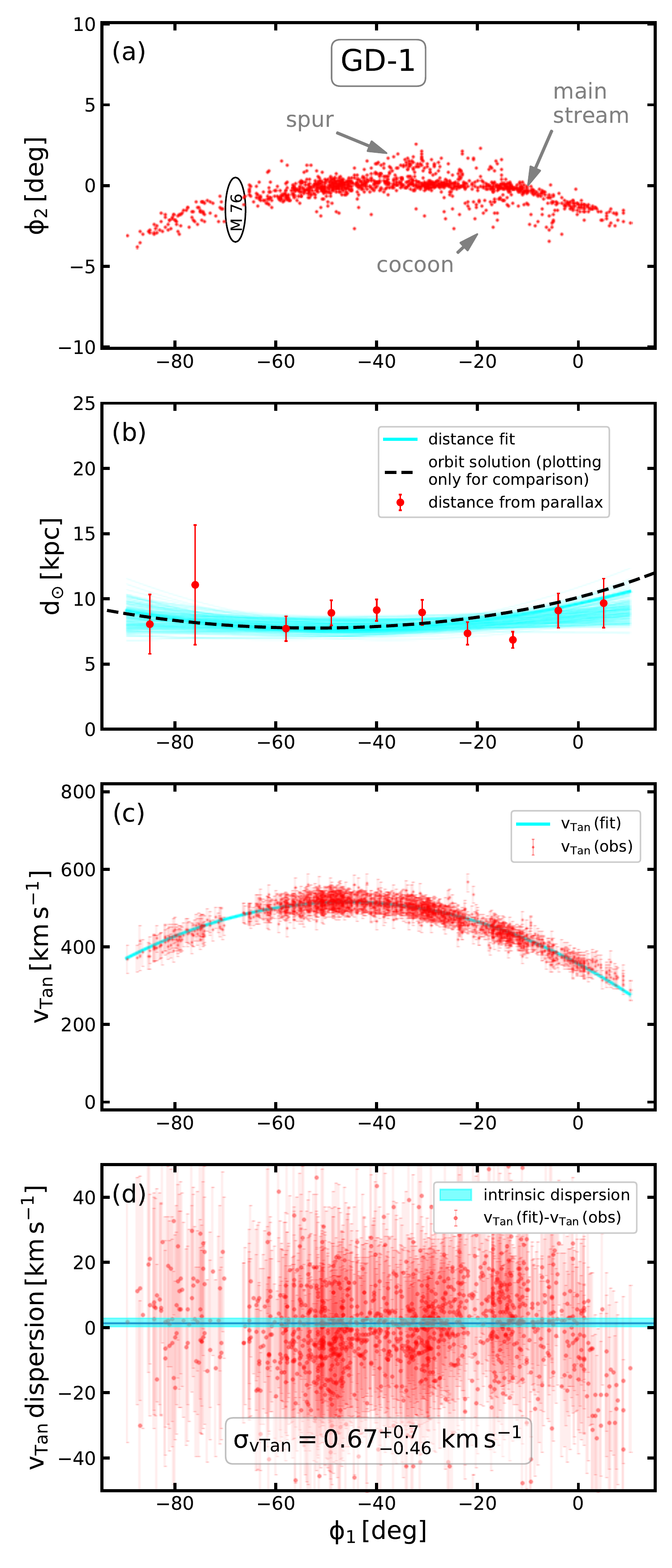}
\includegraphics[width=0.32\hsize]{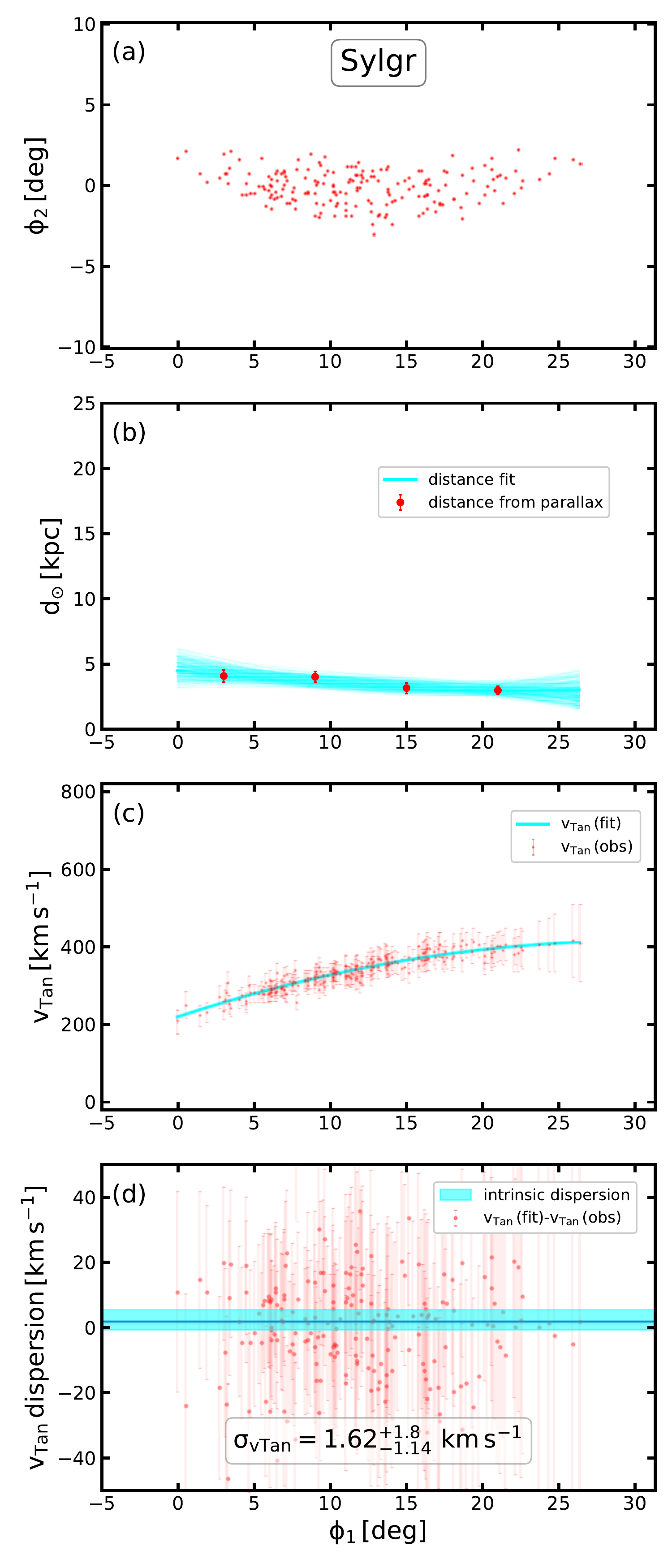}
\includegraphics[width=0.32\hsize]{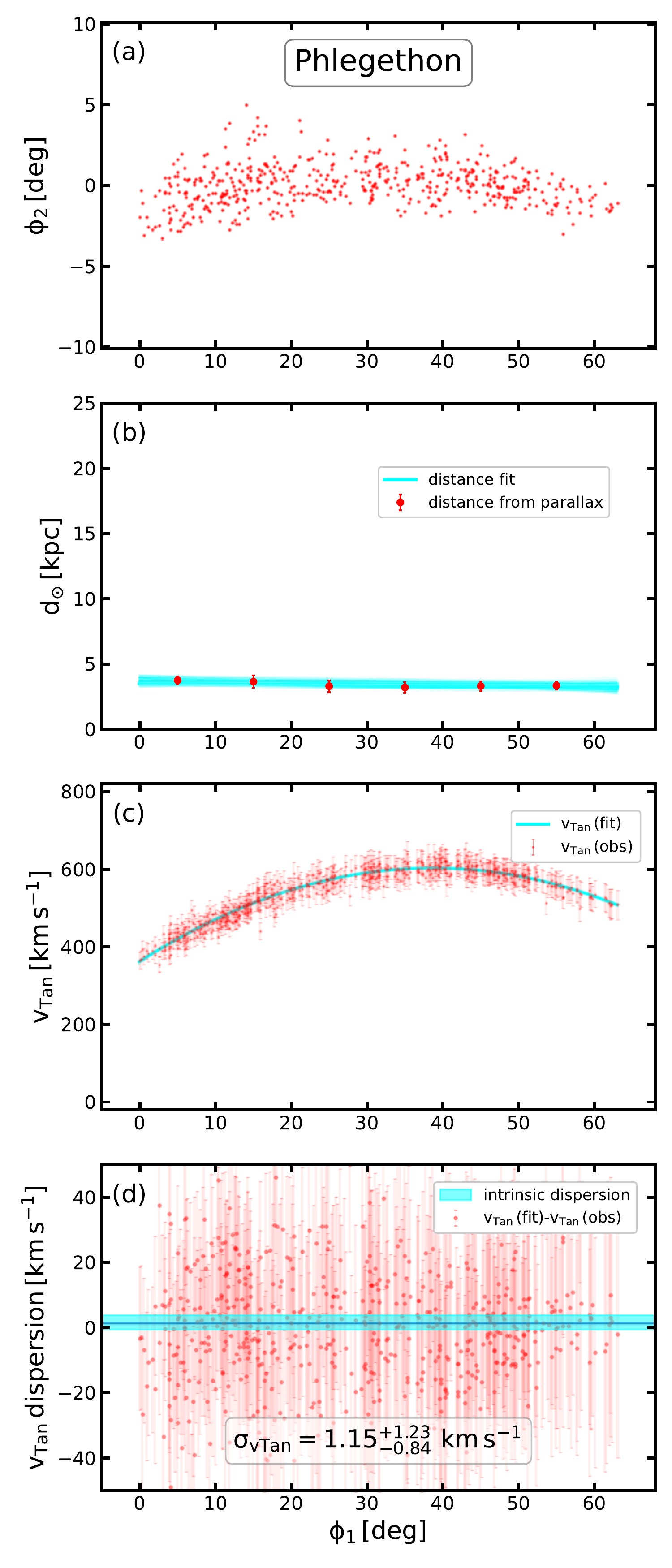}
}
\end{center}
\vspace{-0.5cm}
\caption{Computing tangential velocity dispersion ($\svtan$) of the Milky Way streams using \Gaia\ EDR3. In a given column of this plot, all the panels provide details of a particular stream (the name of the stream is provided in the top panel). In a given column, panel (a) shows positions of the stream stars in the rotated ($\phi_1,\phi_2$) coordinate system, panel (b) shows the distance fit to the stream obtained using \Gaia\ EDR3 parallaxes and photometry (where the fitted curves represent $100$ Monte Carlo representations), and panel (c) shows the $\vtan$ fit to the stream (where ``$v_{\rm Tan}$ (obs)'' is obtained by multiplying distance solutions of panel (b) with the \Gaia\ EDR3 proper motion of stars). Panel (d) shows the residuals of the $v_{\rm Tan}$ (obs) after the mean trend has been subtracted off, and the ``blue band'' represents the intrinsic dispersion. In panel (d), the quoted $\svtan$ value represents the median and the corresponding uncertainties reflect the 16th to 84th percentile range of the distribution (see text). Specifically for ``GD-1'', we compare our distance fit with that of its orbit solution, only to ensure that our distance solutions are reliable (the orbit solution is taken from \citealt{Malhan_2019_Pot}).}
\label{fig:Fig_MW_streams_fit}
\end{figure*}
\begin{figure*}
\begin{center}
\vspace{-0.35cm}
\vbox{
\includegraphics[width=0.32\hsize]{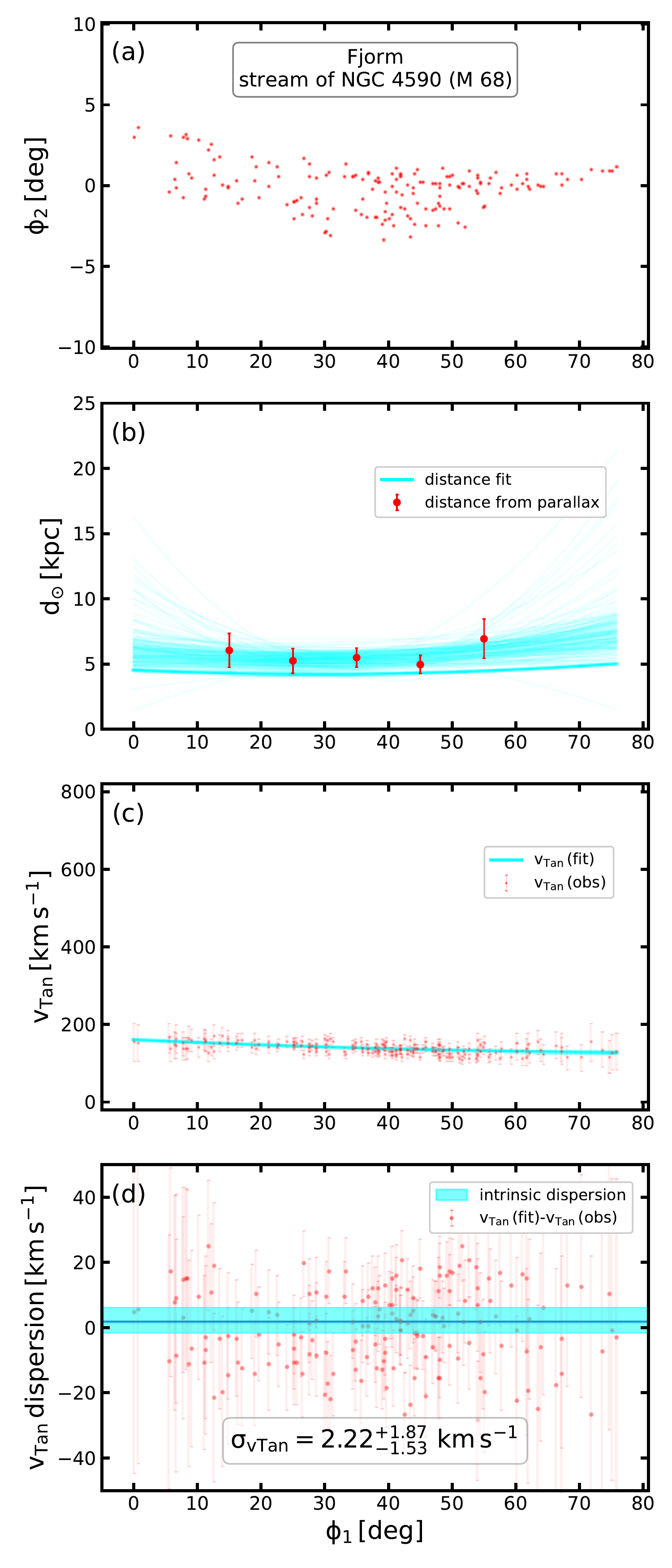}
\includegraphics[width=0.32\hsize]{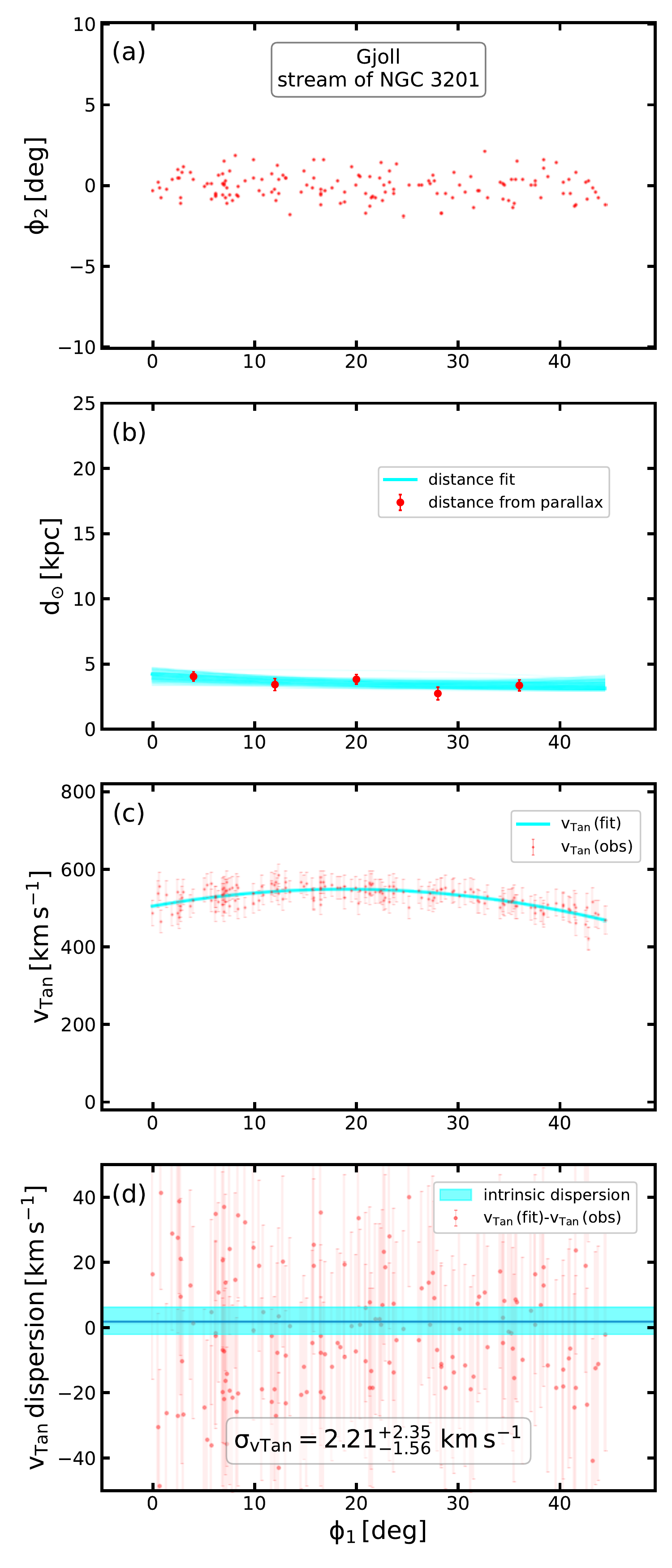}
}
\end{center}
\vspace{-0.5cm}
\caption{Same as Figure~\ref{fig:Fig_MW_streams_fit}, but for different streams.}
\label{fig:Fig_MW_streams_fit2}
\end{figure*}
\section{Tangential velocity dispersions of the Milky Way streams}\label{sec:Tan_vel_of_observed_streams}

There is now mounting evidence that some of the GC streams that orbit the MW halo were accreted from dwarf galaxies (e.g., \citealt{Malhan_cocoon_2019, Malhan_Kshir_2019, Gialluca_2020, Bonaca_2021}). This implies that the $\svtan$ measurement of these streams provide an opportunity to test the prediction that we obtained above, and thus understand whether the parent dwarfs of these streams possessed cuspy or cored DM distribution.

Here, we measure $\svtan$ of $n=5$ streams, namely ``GD-1'', ``Phlegethon'', ``Fj\"orm'', ``Gj\"oll'', ``Sylgr''. The reason for choosing these particular streams is that (1) these are GC streams\footnote{This has been previously established for GD-1 \citep{Malhan_2019_Pot, Bonaca_2020_GD1}, Fj\"orm \citep{Palau_2019}, Gj\"oll \citep{Palau_2021_Gjoll}, and tentatively for Sylgr \citep{Roederer_2019} and Phlegethon \citep{Ibata_2018}.}, (2) these streams have been hypothesized to be of accreted origin (e.g., \citealt{Bonaca_2021, Malhan_2022}), and (3) these are long streams that also possess high stellar densities, and are thus suitable for performing the intended analysis. All of these streams are quite metal poor (with their [Fe/H] lying below $\sim-2$~dex, \citealt{Malhan_2022}), and this further supports the accretion scenario.

The member stars of these streams are taken from the \cite{Ibata_2021_GaiaEDR3} catalogue. The streams in this catalogue were detected in the \Gaia\ EDR3 dataset using the \texttt{STREAMFINDER} algorithm \citep{Malhan_2018_SF, Ibata_2019_GDR2}. In this catalogue, every star possesses \Gaia\ EDR3 based position ($\alpha,\delta$), parallax ($\varpi$), proper motions ($\mu^{*}_{\alpha}, \mu_{\delta}$) and photometry ($G,G_{\rm BP},G_{\rm RP}$), along with the associated uncertainties. 
The photometric information is used along with parallaxes to improve the distance estimates (see below). The parallaxes are corrected for the global parallax zero-point in \Gaia\ EDR3 \citep{GaiaEDR3_Lindegren_2020} and the photometry is corrected for extinction \cite{Ibata_2021_GaiaEDR3}. These streams are shown in Figures~\ref{fig:Fig_MW_streams_fit} and \ref{fig:Fig_MW_streams_fit2}.

To measure $\svtan$ of these streams, we follow a similar procedure as described in Section~\ref{sec:Tan_vel_of_simulations}, with slight modifications in order to account for the observational errors. First we transform the positions of stream stars from ($\alpha,\delta$) to ($\phi_1,\phi_2$) coordinates aligned with each stream. This is shown in panels (a) of Figures~\ref{fig:Fig_MW_streams_fit}, \ref{fig:Fig_MW_streams_fit2}. Next we compute the distance of the stream as a function of $\phi_1$ (which can then be multiplied with proper motions to obtain $\vtan$) (we do not simply compute the average parallax of the stream since this can bias $\svtan$ measurement for streams with distance gradients). Therefore, to properly account for the possible distance gradients, we follow a pragmatic approach. In a given stream, we consider segments along $\phi_1$ of length $\approx10\deg$. This length allows us to have a minimum of $15$ stars in every segment. For each segment we use the stars to compute uncertainty-weighted average mean parallax value (along with the uncertainty on this mean parallax). A reliable estimate of mean parallax value requires high enough number of stars in a given segment. Taking the inverse of this mean parallax provides the average heliocentric distance ($d_{\odot}$) of that segment (along with the uncertainty on $d_{\odot}$). This $d_{\odot}$ value is computed at all the segments of the stream, that provides a means to constrain the distance gradient of the entire stream structure. These distance measurements are shown in panels (b) of Figures~\ref{fig:Fig_MW_streams_fit} and \ref{fig:Fig_MW_streams_fit2}. The typical distance uncertainty (per segment) is $\approx 0.5\kpc$. 

Next, for a given stream, we fit these $d_{\odot}$ measurements using a similar function described by equation~\ref{eq:vtan_fit} (except this time we fit the entire stream at once, and not in individual segments). This fitting is performed using the \texttt{emcee} and it takes into account the uncertainties in $d_{\odot}$ measurements. The posterior on the parameters $a_1,b_1,c_1$ provides the distance fit (as a function of $\phi_1$) and the spread on the posterior provides the uncertainty on this distance fit. Effectively, this procedure allows us to estimate the distance (and the uncertainty) for every star using its $\phi_1$ value. For a given star, we can now multiply its distance with its proper motion to obtain its $\svtan$ (as explained below). 

In passing, we also note that the above distance fitting procedure is augmented with the information on color magnitude diagram (CMD) of stars ([$G_{\rm BP}-G_{\rm RP}, G$], that comes from \Gaia\ EDR3); the CMD information is used as a prior in our likelihood evaluation (see Appendix~\ref{app:CMD}). Since the scatter in the CMDs of all the streams are reduced after this distance correction step, it gives us confidence that the estimated distances are reliable. This is because streams, in general, have distance gradients. Therefore, their observed CMDs are slightly smeared out in apparent magnitude. However, if the observed magnitude of each star is corrected by its ``true'' distance value, then the corrected CMD should have a reduced scatter. Here, we quantify the scatter in a stream's CMD using the k-nearest neighbors algorithm (implemented using \texttt{NearestNeighbors} module in \texttt{sklearn} package). For this, we set the parameter \texttt{n\_neighbors}=10 and \texttt{metric}=euclidean. In Figure~\ref{fig:Fig_MW_streams_CMD}, we compare the distance corrected CMDs with the observed CMDs. Furthermore, we also note that our fitted distance solutions are compatible with the distance measurements of \cite{Bailer-Jones_2021}; as shown in Appendix~\ref{app:dis_vtan}.

In a given stream, to obtain $\vtan$ measurements of the member stars, we multiply the above distance solutions with the \Gaia\ proper motions. For a given star, the uncertainties on the distance solution and on the proper motions provides the uncertainty on the $\vtan$ measurement. Using these $\vtan$ measurements (along with the uncertainties), the stream is fitted using equation~\ref{eq:vtan_fit}; the entire stream structure is fitted at once, and not in segments. The best fit solutions for $\vtan$ for all the streams are shown in panels (c) of Figures~\ref{fig:Fig_MW_streams_fit} and \ref{fig:Fig_MW_streams_fit2}. We further highlight that our fitted $\vtan$ solutions are compatible with the $\vtan$ measurements that one would derive by simply multiplying \cite{Bailer-Jones_2021} distances with \Gaia's proper motions (see Appendix~\ref{app:dis_vtan}).

Finally, to obtain $\svtan$ measurement of a given stream, we subtract-off the above $\vtan$-fit as the systemic velocity of the stream from measured $\vtan$. Then we model the residuals with a Gaussian distribution, including uncertainties on $\vtan$, to derive the $\svtan$ of the stream. These residuals are shown in the panels (d) of the Figures~\ref{fig:Fig_MW_streams_fit} and \ref{fig:Fig_MW_streams_fit2}. For the resulting posterior distribution, its median and $16/84$ percentile provide the $\svtan$ of the stream and the uncertainty on $\svtan$, respectively. These values are shown in panels (e) of Figures~\ref{fig:Fig_MW_streams_fit} and \ref{fig:Fig_MW_streams_fit2} and they are also plotted in Figure~\ref{fig:Fig_Summary}. 

In Appendix~\ref{app:robust} we demonstrate that these $\svtan$ measurements of the streams are robust.

Table~\ref{tab:table_streams} provides the $z$-score (for two-tailed hypothesis test) and the corresponding $p$-value for the null hypothesis that an observed $\svtan$ measurement (with its associated uncertainty) is drawn from the Gaussian distribution for one of the 5 simulation scenarios shown in the lower panel of Figure~~\ref{fig:Fig_Summary}. For a given stream $s$ and a given scenario $i$, the z-score is computed as:
\begin{equation}\label{eq:likelihood}
\begin{aligned}
    z &= (\svtan^s -\svtan^i)/\sigma\,,
\end{aligned}
\end{equation}
where $\svtan^s$ is the $\svtan$ measurement of the observed stream $s$, $\svtan^i$ corresponds to that of the scenario $i$, and $\sigma$ is the sum in quadrature of the uncertainties on these two quantities. A given $p$-value implies that the probability that the observed stream was drawn from the population describing a given simulation scenario can be rejected with confidence of $(1-p)\times 100$\% (e.g. $p=0.01$, implies the null hypothesis can be rejected at the 99\% level).

\begin{table*}
\caption{z-scores (and p-values) for Milky Way stream streams being drawn from various simulated stream populations}
\label{tab:table_streams}
\begin{centering}
\begin{tabular}{|l|c|c|c|c|c|}
\hline
\hline
MW stream & in-situ & SCu & LCu & SCo & LCo\\
\hline
\hline
& & & & &\\

GD-1  &  0.361  ($ 0.718 $)  &  -4.162  ($ <10^{-3} $)  &  -9.712  ($ <10^{-3} $)  &  -1.221  ($ 0.222 $)  &  -1.172  ($ 0.241 $)\\

Sylgr  &  0.979  ($ 0.328 $)  &  -1.148  ($ 0.251 $)  &  -3.731  ($ <10^{-3} $)  &  0.017  ($ 0.986 $)  &  -0.233  ($ 0.816 $)\\

Phlegethon  &  0.768  ($ 0.442 $)  &  -2.046  ($ 0.041 $)  &  -5.683  ($ <10^{-3} $)  &  -0.355  ($ 0.722 $)  &  -0.599  ($ 0.549 $)\\

Fjorm  &  1.122  ($ 0.262 $)  &  -0.787  ($ 0.431 $)  &  -3.284  ($ 0.001 $)  &  0.398  ($ 0.691 $)  &  0.066  ($ 0.948 $)\\

Gjoll  &  1.094  ($ 0.274 $)  &  -0.632  ($ 0.528 $)  &  -2.639  ($ 0.008 $)  &  0.384  ($ 0.701 $)  &  0.059  ($ 0.953 $)\\

& & & & & \\
\hline
\hline
\end{tabular}
\end{centering}
\tablecomments{Left most column provides the name of the observed stream; the next 4 columns give $z$-score (and corresponding $p$-values) for the hypothesis that the observed stream is drawn from the Gaussian distribution describing simulated streams from: the in-situ scenario, the SCu dwarf galaxy scenario, the LCu scenario, the SCo scenario and the LCo scenario.}
\end{table*}
\section{Conclusion and Discussion}\label{sec:Conclusion}

We draw our main conclusions by inspecting Figure~\ref{fig:Fig_Summary} and Table~\ref{tab:table_streams}. They compare the predicted values of $\svtan$ (that we obtained by analysing N-body GC stream models produced in different DM scenarios) with the observations (coming from the MW streams). The bottom panel of Figure~\ref{fig:Fig_Summary} shows Gaussians that quantify the scatter in $\svtan$ measurements of the simulated streams produced in in situ/cored/cuspy scenarios. These Gaussians imply that: 1) in situ GC streams should possess $\langle \svtan \rangle = 0.5\pm0.2\kms$, 2) GC streams accreted inside the cuspy SCu (LCu) subhalo with mass $M_0= 10^8(10^9)\msun$ should possess $\langle \svtan \rangle = 3.7\pm0.2 \, (8.5\pm0.4)\kms$, and 3) GC streams accreted inside the cored SCo (LCo) subhalo with mass $M_0= 10^8(10^9)\msun$ should possess $\langle \svtan \rangle= 1.6\pm0.3 \, (2.1\pm 1.0)\kms$. We summarize our main results below.

\begin{itemize}
\item  N-body simulations of GC tidal streams accreted from dwarf galaxies with different central DM density profiles (cuspy vs. cored) show that there are significant and measurable differences in the observed $\svtan$ (the tangential velocity dispersion stars in the stream) that reflect the nature of the central density profiles of their parent dwarf galaxies. 

\item Current \Gaia\ EDR3 proper motions and parallaxes are used to determine $\svtan$ for 5 GC streams  (``GD-1'', ``Phlegethon'', ``Fj\"orm'', ``Gj\"oll'', ``Sylgr'') studied in this work. It is not possible with current \Gaia\ observational uncertainties to reject the hypothesis that these streams were formed in situ. Most of the observed GC streams in this study orbit at a galactocentric distance of $\approx 20\kpc$, while the in situ stream models in \citep{Malhan_DM_2021} were simulated with orbital radii of $60\kpc$. Therefore, for a fair comparison, we ran $5$ additional N-body simulations of streams under the in situ framework, but this time adopting the GC's orbital radius as $\approx 20\kpc$. These additional streams can be seen as the top $5$ red markers in Figure~\ref{fig:Fig_Summary}. These additional simulations do not alter our conclusion on this point. 

\item If however, the progenitor GCs of the MW streams analysed here were indeed accreted as previously argued (see below), our $\svtan$ measurements enable us to reject with high confidence the hypothesis that their parent dwarf galaxies were cuspy with $M_0\simgt10^9\msun$. We can also reject higher mass cuspy subhalos since  GC streams from such dwarfs are expected be even hotter. Also, it is not possible that these MW streams would have originated from lower-mass cuspy subhalos because dwarfs with $M_0\simlt10^8\msun$ are not expected to host any GC population (e.g., \citealt{Forbes_2018}). In view of these arguments, our current analysis disfavours the cuspy CDM subhalos. 

\item The \Gaia\ uncertainties on proper motions and parallaxes are currently too large to definitively determine whether the parent subhalos of these streams were cored or cuspy with $M_0=10^8\msun$.

\item Additionally, we recompute $\svtan$ of 5 GC streams by incorporating the ``systematic errors'' present in \Gaia\ EDR3's proper motions and parallaxes (for details on these errors, see \citealt{GaiaEDR3_Lindegren_2020}). This analysis is performed to examine the impact of these errors on the $\svtan$ values that we measure in this work. As shown in Appendix~\ref{app:systematic_errors}, the inclusion of these systematic errors only minutely change the $\svtan$ measurements, and they do not affect the final conclusion of this work.

\end{itemize}

Although we are unable to definitively rule out (based on the kinematic analysis done here) the possibility that the progenitors of these streams were in situ GCs, there are other lines of evidence that indicate most of them have an accreted origin. The in situ GC population is overall redder and more metal rich than the accereted GC population \citep[e.g.][]{Kruijssen_etal_2019}. Furthermore, orbital action space clustering of GCs and halo stars and a comparison of the metallicities of GCs and those same halo stars has been used to assign many accreted GCs, including GD1 to previous merger events \citep{myeong_vasiliev_19,Massari_etal_2019, Kruijssen_2020_kraken, Bonaca_2021, Malhan_2022}. The metal rich in situ GC population has a slight net prograde rotation, while the accreted GC population has no net rotation but subsets associated with specific accretion events can be seen to be clustered in angular-momentum \citep{Massari_etal_2019}. In addition to having a nearly circular and retrograde orbit, GD1 is extremely metal poor with a mean metallicity of $-2.2$~dex \citep{Malhan_2019_Pot}, much closer to the metallicity of dwarf spheroidal satellites of the MW \citep{Kirby_2013} than in situ GCs \citep{Zinn_1985}. 

In addition to $\svtan$, the other two stream parameters that are also useful to probe the DM density profiles inside dwarfs are: transverse physical widths ($w$) and dispersion in the los velocity ($\sv$).  \citet{Malhan_DM_2021} showed that in situ GCs produce streams with $(\langle w \rangle, \langle \sv \rangle)=(45\pm 15\pc, 0.7\pm 0.2\kms)$, GC streams accreted in cuspy subhalos produce  with $(\langle w \rangle, \langle \sv \rangle)\simgt(650\pc,4\kms)$, and somewhat smaller widths $(\langle w \rangle, \langle \sv \rangle)\sim(90-500\pc,<4\kms)$ result when GCs accrete inside cored subhalos\footnote{These constrains are based on the subhalo models with mass $M_0=10^8\msun,10^9\msun$}. A combination of multiple parameters could provide a stronger means to probe the DM density profile inside the parent dwarf. For instance, we can in principle compare the predicted $w$ values with the recent $w$ measurements of other MW streams (e.g., \citealt{Bonaca_2020, Ferguson_2022AJ, Tavangar_2022}) to comment on their ``accretion'' origin.  For GD-1, while its $\svtan$ measurement appears to be more consistent with the in-situ scenario (see Table~\ref{tab:table_streams}), consideration of these additional parameters: $w$ ($=130^{+30}_{-20}$~pc, \citealt{Malhan_cocoon_2019}) and $\sv$ ($=2.1\pm0.3\kms$, \citealt{Gialluca_2020}) suggest that GD-1 was likely accreted inside a cored subhalo. 

In summary, our analysis indicates that 4 (out of 5) MW streams shown in Figure~\ref{fig:Fig_Summary} favor cored DM subhalos over cuspy CDM subhalos. Although this inference is based on only two subhalo masses (i.e., $M_0=10^8\msun$ and $10^9\msun$), we argue that it is unlikely that these streams could have accreted inside cuspy subhalos of higher mass since such streams would be even hotter.

The origin of cored subhalos is still hotly debated. While cored subhalos are favored by alternative DM candidates, hydrodynamical simulations have shown that DM cores can result from erasure of DM cusps if the dwarf galaxy had a sufficiently vigrous and episodic star formation phase (e.g. \citealt{Pontzen2012}). Under such a scenario, the resulting cored subhalo would still be consistent with the CDM paradigm. Recent cosmological hydrodynamic simulations predict that subhalos with $M_0\simlt 10^{10}\msun$ would have formed too few stars over their lifetimes, and the resulting baryonic feedback is too weak to unbind their DM cusps (e.g., \citealt{Lazar_2020}). If a significant fraction of  tidal streams from accreted GCs are found to be dynamically consistent with having originated from cored subhalos with $M_0\simlt 10^{10}\msun$, then we may be forced to move to models beyond CDM. However additional simulations with a greater variety of dwarf galaxy properties and orbital initial conditions are needed before firm conclusions can be drawn.

Cosmological hydrodynamical zoom-in simulations with different types of dark matter: CDM, WDM, SIDM and mixed DM, e.g. WDM with self-interaction, \citep{Fitts_etal_2019} show that the addition of baryons substantially decrease differences between the simulations with different types of DM. However baryons decrease the sizes of cores in SIDM and WDM+SIDM subhalos compared to SIDM-only simulations, but they have significantly lower central densities than CDM-only halos.  In future, it will be interesting to simulate a wider variety of cored subhalo models (by varying their mass ranges, physical sizes, core sizes and inner density slopes). 

In \citet{Malhan_DM_2021}, we showed that three observationally determinable quantities for accreted GC streams: physical width $w$, line-of-sight velocity dispersion $\sigma_{\mathrm vlos}$ and dispersion in the $z$-component of angular momentum $L_z$, were all sensitive probes of the degree of tidal heating experienced by a GC stream in its parent dwarf galaxy and could enable us to set constraints on the DM profiles of dwarf galaxies. In this work we have shown, in addition, that $\svtan$ is able to provide similar discrimination.

In the future, we will consider additional heating arising from passage of the stream through the disk or interactions with molecular clouds \citep{Amorisco_2016} or the bar \citep{Pearson_2017}. Furthermore,we will assess whether all 6 phase space coordinates when combined may yield stronger constraints on DM. In practice however stream membership is difficult to assess in the absence of spectroscopy and accurate \Gaia\ proper motions, especially for distant streams. While radial velocities have the smallest uncertainties -- e.g.  $1-2\kms$ uncertainty for $G\lesssim 19$ with current large multiobject spectrographs like DESI \citep{DESI_2016a, DESI_2016b, DESI-MWS_RNAAS_2020,DESI_MWS_2022} -- the fact that tidal streams generally extend over tens of degrees on the sky make it extremely expensive observationally to obtain the large numbers of
$\vlos$ measurements needed to reliably compute $\sv$ for many streams. While \Gaia\ DR3 released  $\vlos$ for over 30 million stars brighter than $G=14$,  Figure~4 shows that most of the stars of interest here are fainter than this magnitude limit.

The metric we study in this work, $\svtan$, depends on accurate measurements of both the proper motions of stream stars and their distances. We obtained both quantities in this work from \Gaia\ EDR3 observations. Future \Gaia\ data releases are expected to decrease the uncertainties on both the measured proper motions and parallaxes by around 50\% for each quantity relative to EDR3 uncertainties \citep[see,][and the Gaia-ESA website\footnote{https://www.cosmos.esa.int/web/gaia/science-performance}]{Gaia_EDR3_basic} resulting in a net decrease in the uncertainty on $\svtan$ of $\sim 60-65$\% for the streams we consider here.  If both $\sv$ and $\svtan$ are available for a significant sample of stars, one might combine them to obtain a 3D velocity dispersion, but currently adequate numbers of 
$v_{\mathrm los}$  measurements do not exist for the streams considered here. At the present time and for the foreseeable future, \Gaia\ proper motions and parallaxes, being the most abundantly measured quantities, offer the best way to quantify the velocity dispersions of GC tidal streams.

\section*{ACKNOWLEDGEMENTS}

We thank our referee for their constructive and valuable comments. KM and KF acknowledge support  from the $\rm{Vetenskapsr\mathring{a}de}$t (Swedish Research Council) through contract No. 638-2013-8993 and the Oskar Klein Centre for Cosmoparticle Physics. KM acknowledges support from the Alexander von Humboldt Foundation at Max-Planck-Institut f\"ur Astronomie, Heidelberg. KM is also grateful to the IAU's Gruber Foundation Fellowship Programme for their finanacial support. MV is supported by NASA-ATP award 80NSSC20K0509. KF gratefully acknowledges support from the Jeff and Gail Kodosky Endowed Chair in Physics at the University of Texas, Austin; the U.S. Department of Energy, Office of Science, Office of High Energy Physics program under Award Number DE-SC0022021 at the University of Texas, Austin; the DoE grant DE- SC007859 at the University of Michigan; and the Leinweber Center for Theoretical Physics at the University of Michigan. RI acknowledges funding from the European Research Council (ERC) under the European Unions Horizon 2020 research and innovation programme (grant agreement No. 834148). 

This work has made use of data from the European Space Agency (ESA) mission
{\it Gaia} (\url{https://www.cosmos.esa.int/gaia}), processed by the {\it Gaia}
Data Processing and Analysis Consortium (DPAC,
\url{https://www.cosmos.esa.int/web/gaia/dpac/consortium}). Funding for the DPAC
has been provided by national institutions, in particular the institutions
participating in the {\it Gaia} Multilateral Agreement.

\bibliography{ref1}
\bibliographystyle{aasjournal}

\appendix

\section{Comparing the observed and distance-corrected CMDs of streams}\label{app:CMD}

In Section~\ref{sec:Tan_vel_of_observed_streams}, we perform distance fitting to the streams as a function of their $\phi_1$ coordinate. For this distance fitting, we use \Gaia's parallaxes and also \Gaia's photometry ($G_{\rm BP}-G_{\rm RP}, G$). The reason for using the photometry information can be explained as follows. Stellar streams generally possess distance gradients along their lengths, and therefore their observed CMDs are slightly smeared out in apparent magnitude (here, $G$ magnitude). However, if the photometry of each star is corrected by its ``true'' distance value, then the corrected CMD will have less scatter. Therefore, this additional information on the ``CMD scatter'' provides a means to better constrain the streams' distance solutions. For this, during our distance fitting procedure, we impose a (constant) prior condition in likelihood evaluation that -- the resulting distance solution should be such that produces a CMD with less scatter than the observed CMD.

\begin{figure*}
\begin{center}
\vspace{-0.3cm}
\hbox{
\includegraphics[width=0.32\hsize]{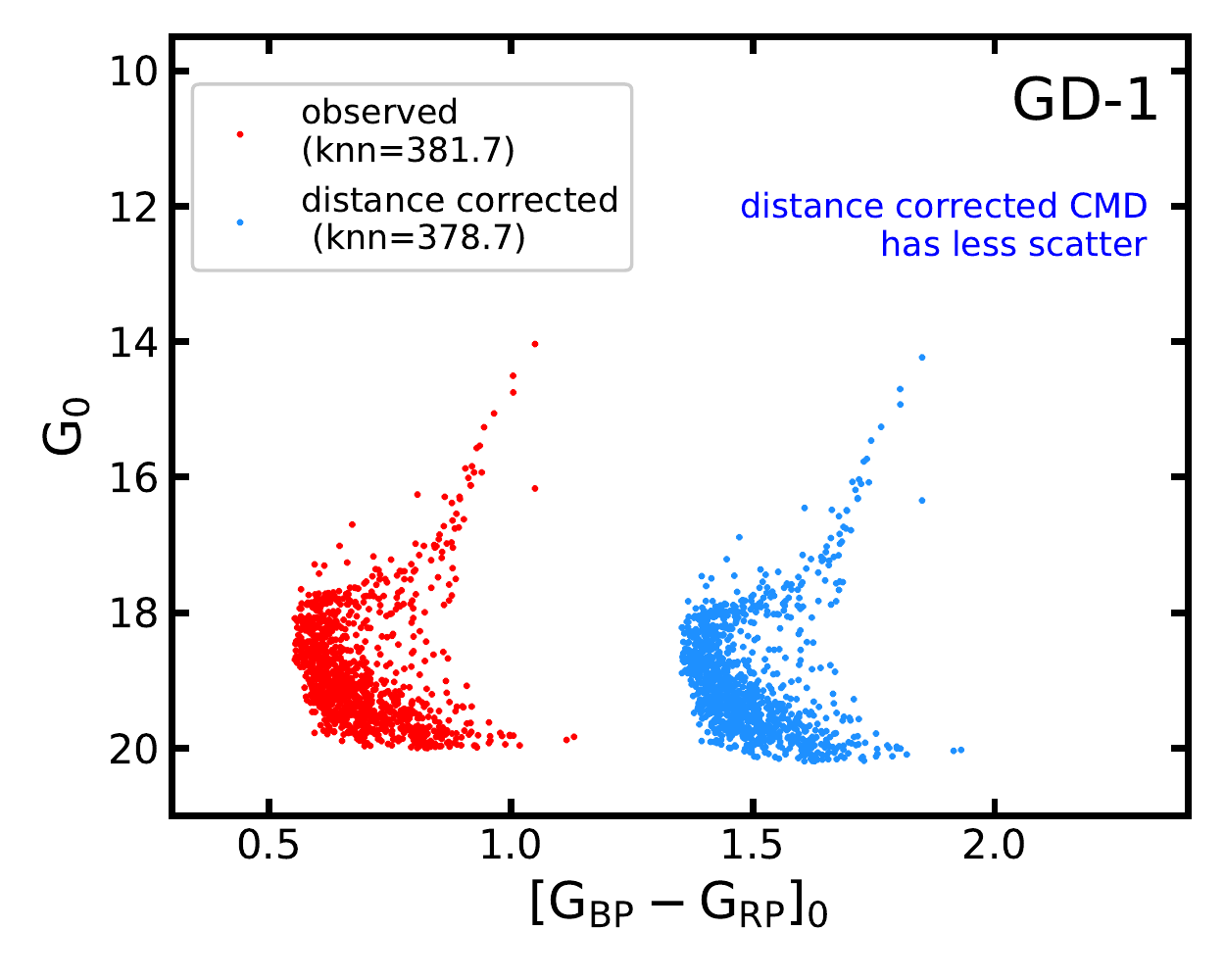}
\includegraphics[width=0.32\hsize]{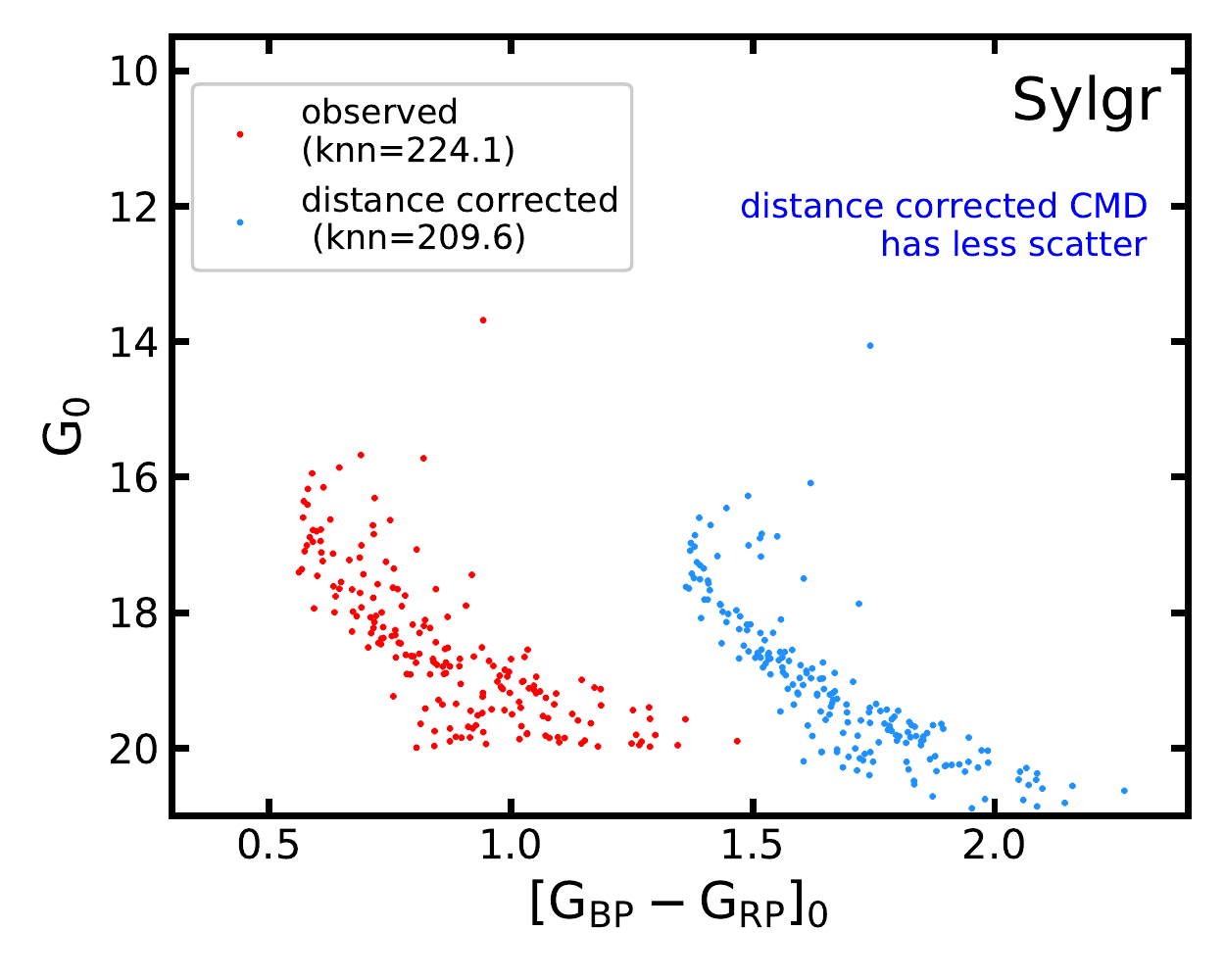}
\includegraphics[width=0.32\hsize]{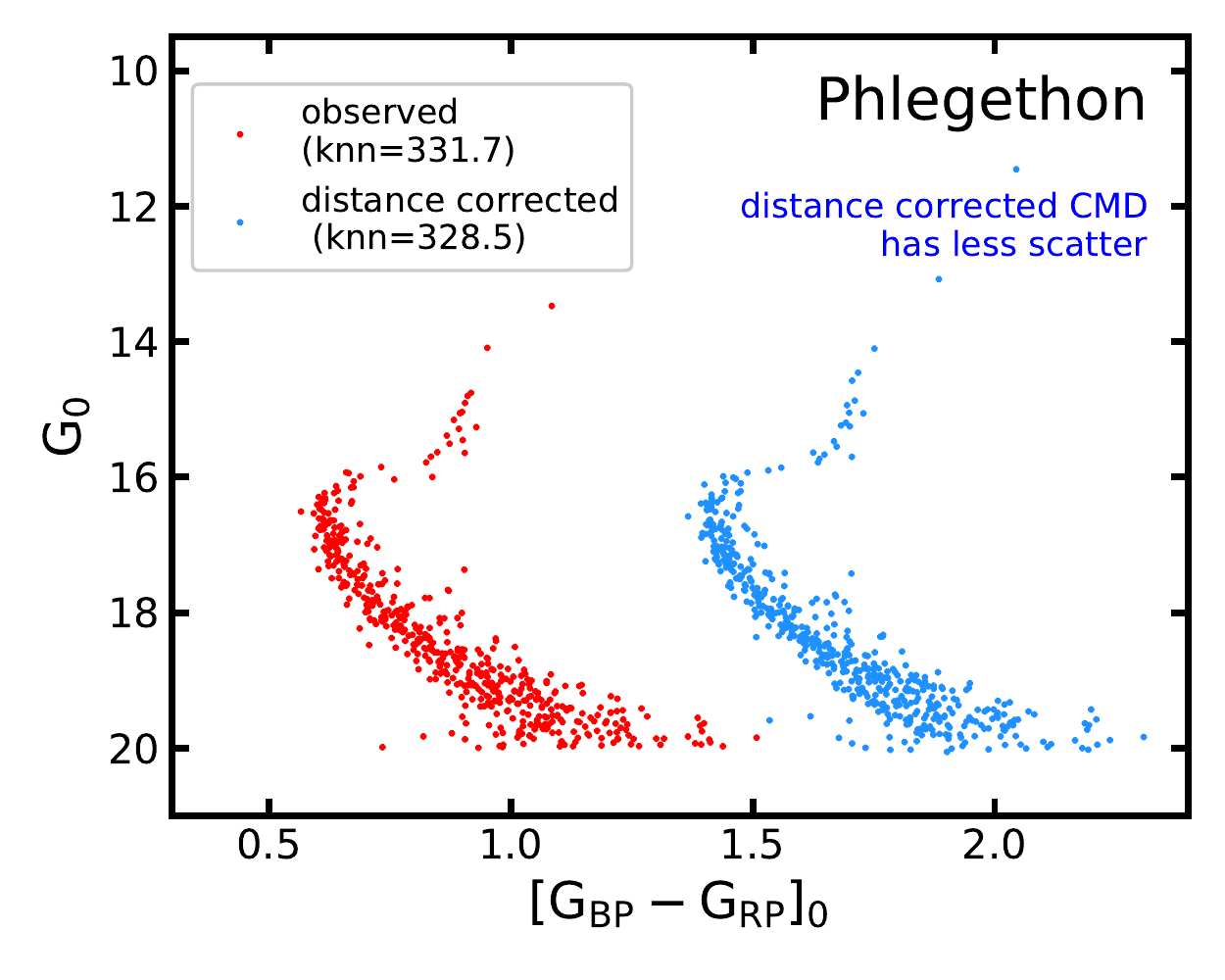}
}
\hbox{
\includegraphics[width=0.32\hsize]{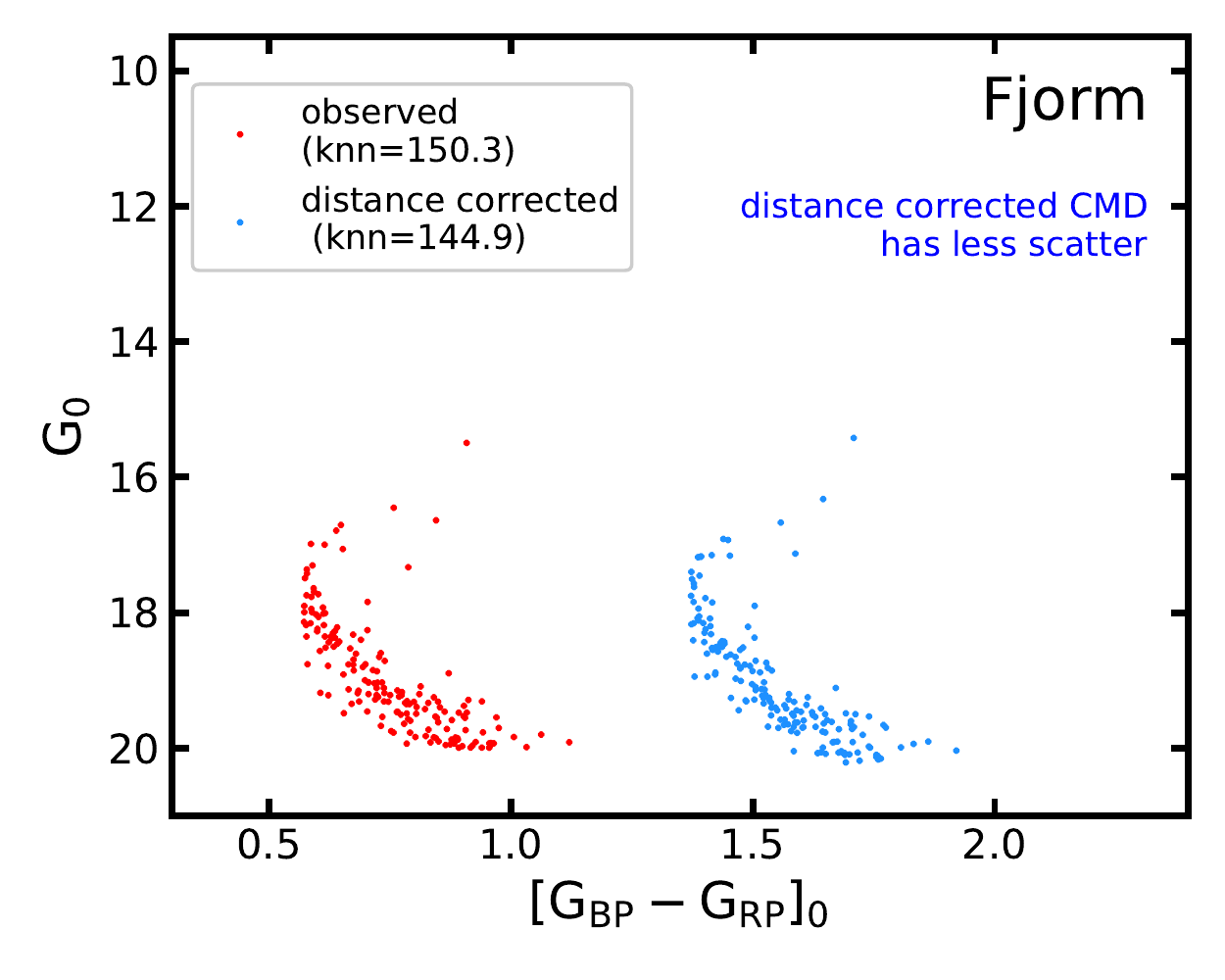}
\includegraphics[width=0.32\hsize]{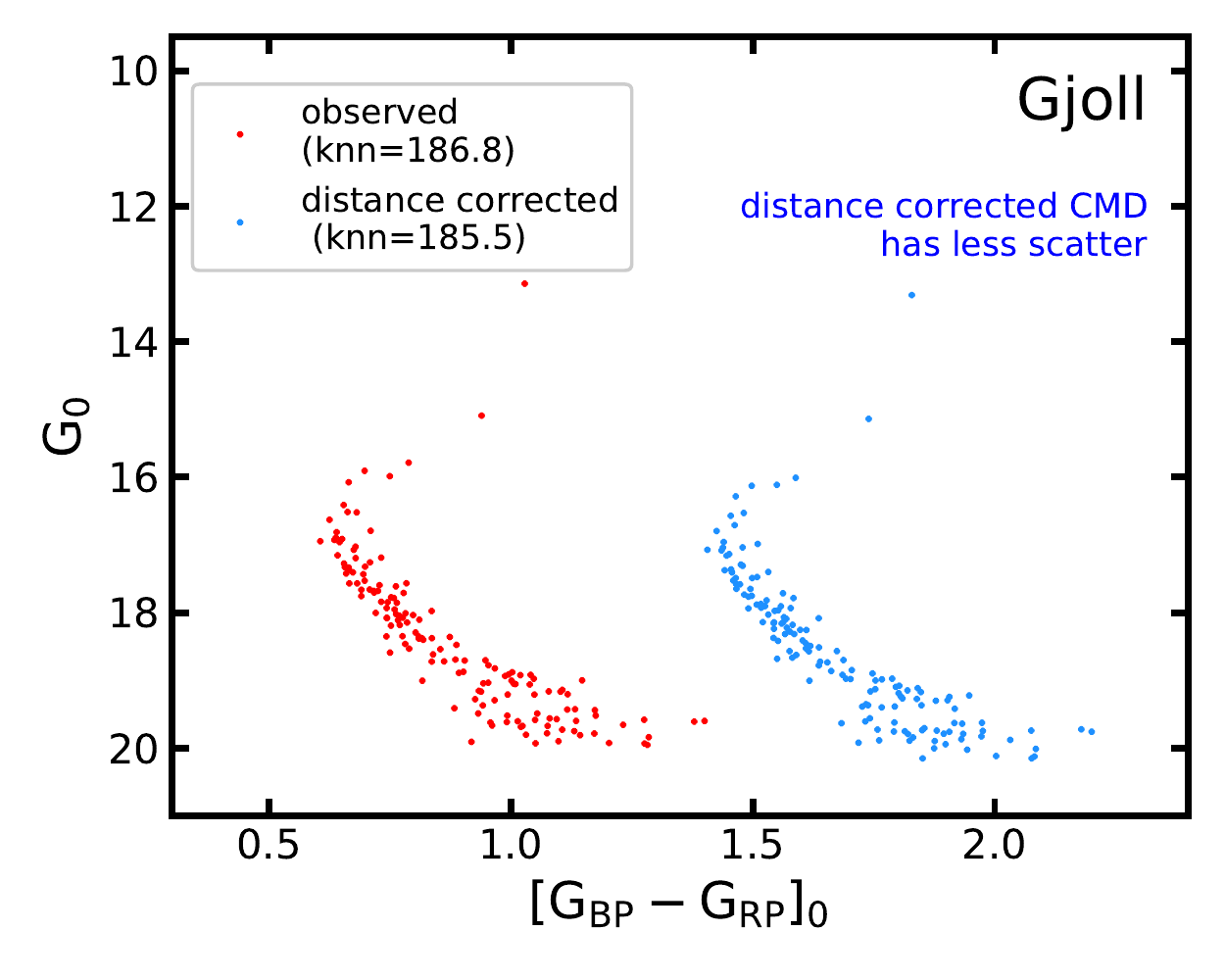}
}
\end{center}
\vspace{-0.5cm}
\caption{Color Magnitude Diagrams (CMDs) for 5 observed stellar streams (as indicated by labels) constructed from \Gaia\ photometry. In each panel the observed, extinction corrected CMD (left, red points) and distance corrected  CMD (right, blue points) are shown. The $knn$ values in the legend of each panel estimate the scatter in each CMD (smaller $knn$ values imply less scatter).}
\label{fig:Fig_MW_streams_CMD}
\end{figure*}

The corresponding result is shown in Figure~\ref{fig:Fig_MW_streams_CMD}, that compares the ``observed'' and ``distance corrected'' CMDs of all the streams. The scatter in these CMDs are quantified using the \texttt{NearestNeighbors} module, and this confirms that the distance corrected CMDs have less scatter than the observed CMDs. This result can also be discerned by visually inspecting Figure~\ref{fig:Fig_MW_streams_CMD}. This implies that our distance solutions are reliable.

\section{Examining the accuracy of our fitted solutions for $d_{\odot}$ and $\vtan$}\label{app:dis_vtan}

We assess the reliability of our fitted solutions for the distances ($d_{\odot}$) and the tangential velocities ($\vtan$) of the stream stars (shown in panels b and d of Figures~\ref{fig:Fig_MW_streams_fit} and \ref{fig:Fig_MW_streams_fit2}) as follows.

\begin{figure}
\begin{center}
\vspace{-0.3cm}
\includegraphics[width=\hsize]{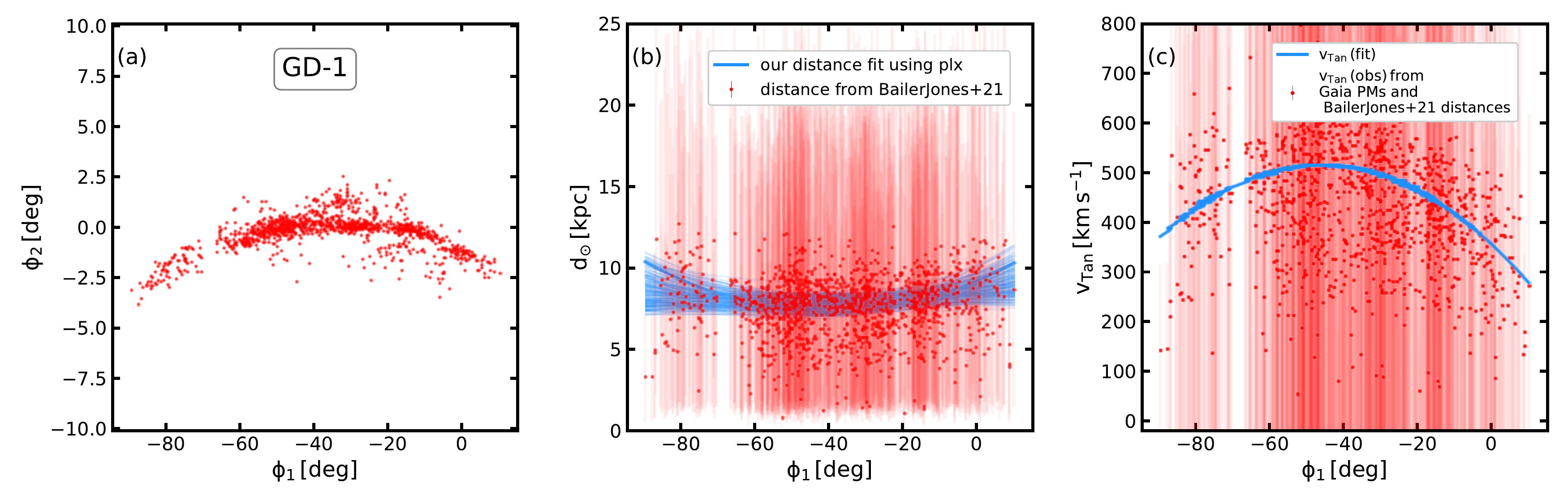}
\end{center}
\vspace{-0.5cm}
\caption{Comparing our fitted $d_{\odot}$ and $\vtan$ solutions with those obtained from \cite{Bailer-Jones_2021} and \Gaia\ EDR3. This plot corresponds to the ``GD-1'' stream.}
\label{fig:Fig_dis_compare}
\end{figure}

We compare our fitted $d_{\odot}$ solutions (based on the \Gaia\ EDR3 parallaxes, see main text) with the $d_{\odot}$ measurements from the \cite{Bailer-Jones_2021} catalogue. This comparison is shown in Figure~\ref{fig:Fig_dis_compare}b for the GD-1 stream. Based on the visual inspection, we conclude that our solutions are consistent with these measurements. We also note that uncertainties on distances of the individuals stars from \cite{Bailer-Jones_2021} are very large ($\sim8\kpc$). We repeated this exercise for other streams as well and found similar consistency. 

Finally, we compare our fitted $\vtan$ solutions with those derived by simply multiplying $d_{\odot}$ from \cite{Bailer-Jones_2021} and proper motions from \Gaia\ EDR3. This comparison is shown in Figure~\ref{fig:Fig_dis_compare}c for GD-1. Based on the visual inspection, we conclude that our solutions are consistent with these measurements; although the uncertainties on the measurements of the individual stars are very large ($\sim550\kms$). We repeated this exercise for other streams as well and found similar consistency, leading us to conclude that our fitted $d_{\odot}$ and $\vtan$ solutions are reliable.

\section{Examining the robustness of the measured $\svtan$ of the Milky Way streams}\label{app:robust}

\subsection{Examining the robustness due to possible mis-estimates of the observational uncertainties on $\vtan$}

For the Milky Way streams, we note that their member stars possess quite large observational uncertainties on $\vtan$ (of the order of $\sim20\kms$, see panels `d' in Figures~\ref{fig:Fig_MW_streams_fit} and \ref{fig:Fig_MW_streams_fit2}). However, we constrain the $\svtan$ to the order of $\sim 1-2\kms$. In this appendix we demonstrate that 
even though the $\vtan$ uncertainties on tangential velocity measurements for individual stars are large our method is sensitive to the changes in these large uncertainties and able to measure intrinsic velocity dispersions that are much smaller than the current uncertainties. 

To illustrate this we take the Phlegethon stream and artificially modify the $\vtan$ uncertainties on individual stars and recompute the $\svtan$, always keeping the $\vtan$ measurements unchanged (i.e. only modifying the uncertainties). In the first case, we set these uncertainties to $0\kms$ which results in $\svtan=17.19\kms$ by applying equivalent of eq.~\ref{eq:likelihood} (see Section~\ref{sec:Tan_vel_of_observed_streams}). This value is much larger than the value mentioned above for Phlegethon, but this is expected because now the uncertainty-term (in eq.~\ref{eq:likelihood}) attributes the entire spread in the ``observed $\vtan$ (obs) – $\vtan$ (fit)'' distribution (i.e., residuals shown in panels `d' in Figure~\ref{fig:Fig_MW_streams_fit}) to the internal dispersion of the stream $\svtan$. In the second case, we decrease the $\vtan$ uncertainty to half of the actual values and measure $\svtan=12.25\kms$. Note that this value is smaller than the one computed in the first case because now the  spread in the residual distribution is being shared by the uncertainty-term (that are finite and non-zero) and the internal dispersion $\svtan$. In the third case we decrease the uncertainties to $80\%$ of the actual values and measure $\svtan= 1.29\kms$. As expected $\svtan$ decreases further because  now the velocity uncertainties absorb a larger share of the residual distribution. This explains both why the measured intrinsic dispersion is so much smaller than the observed dispersion and why we assert that a decrease in the uncertainties on $\vtan$ expected from future Gaia data releases will improve these $\svtan$ measurements.

\subsection{Determining the effects of correlations}

We assess the effects of correlations between uncertainties in proper motions and parallax in the following way. First, we take the Phlegethon stream and shuffle the proper motion uncertainties of its stars while keeping the parallax uncertainties unchanged (i.e., we randomly reassign the proper motion uncertainty of star $j$ to star $k$ and star $k$ to $i$, and so on). We do this $10$ times to examine whether the resulting $\svtan$ (on an average) is same as what we report above. Based on this we find that $\svtan$ (on an average) changes by only $+1.5 \%$. We repeat the above exercise, except this time we shuffle the parallax uncertainties between stars while keeping the proper motion uncertainties unchanged. In this case we find that the resulting $\svtan$ (on an average) changes by only $-2.7\%$. Finally, we repeat the above exercise with a few other streams and find similarly small changes in our estimated $\svtan$ measurements. This suggests that the correlations should have minor effects on the reported $\svtan$ values of the streams.

\section{Examining the impact of systematic errors on the measured $\svtan$ of the Milky Way streams}\label{app:systematic_errors}

We recompute $\svtan$ of 5 GC streams by incorporating the ``systematic errors'' present in Gaia EDR3's proper motions and parallaxes. These errors are provided in Section~5.6 of \cite{GaiaEDR3_Lindegren_2020} as $0.0108$~mas in $\varpi$, $0.0112\masyr$ in $\mu^{*}_{\alpha}$ and and $0.0107\masyr$ in $\mu_{\delta}$. These values essentially put a floor on the precision with which parallaxes and proper motions are measured.

To recompute $\svtan$, we do the following. For a given stream, we consider the individual stars, and to these we add the above errors (in quadrature) to the observed \Gaia\ uncertainties in parallaxes and proper motions. This essentially inflates the uncertainties of every star. Then, we compute $\svtan$ by following the same procedure as described in Section~\ref{sec:Tan_vel_of_observed_streams}. The final $\svtan$ values are provided in Table~\ref{tab:table_systematic}. Table~\ref{tab:table_systematic} also provides the $p-$values for the null hypothesis that these new $\svtan$ values (with their associated uncertainties) are drawn from the counterpart $\svtan$ measurements that we computed in Section~\ref{sec:Tan_vel_of_observed_streams} without the inclusion of systematic errors. To compute these $p-$values, we follow the same method described in Section~\ref{sec:Tan_vel_of_observed_streams}. The fact that these $p-$values are $\sim 1$ indicate that, for a given stream, the two types of $\svtan$ measurements are similar.

Table~\ref{tab:table_systematic_r} is similar to Table~\ref{tab:table_streams}, except this time produced using the new $\svtan$ measurements. The fact that values in Table~\ref{tab:table_systematic_r} are qualitatively similar to those present in Table~\ref{tab:table_streams} suggests that inclusion of systematic errors do not affect our final conclusion in regard to the cusp/core scenario of the parent subhalos.

\begin{table*}
\caption{$\svtan$ of Milky Way streams (in $\masyr$) computed by including the systematic errors. The values in the brackets provide the $p-$values of these new $\svtan$ measurements being drawn from their counterpart streams whose $\svtan$ were measured without including the systematic errors.}
\label{tab:table_systematic}
\begin{centering}
\begin{tabular}{|l|c|c|c|c|}
\hline
\hline
GD-1 & Sylgr & Phlegethon & Fj\"orm & Gj\"oll\\
\hline
& & & &\\

$ 0.63 ^{+ 0.67 }_ {- 0.45 }\,( 0.963 )$  &  $ 1.76 ^{+ 1.83 }_ {- 1.23 }\,( 0.95 )$
  &  $ 1.12 ^{+ 1.23 }_ {- 0.78 }\,( 0.984 )$  &  $ 1.86 ^{+ 1.66 }_ {- 1.29 }\,( 0.874 )$  
 &  $ 2.19 ^{+ 2.25 }_ {- 1.53 }\,( 0.994 )$ \\

& & & & \\
\hline
\hline
\end{tabular}
\end{centering}
\end{table*}

\begin{table*}
\caption{Same as Table~\ref{tab:table_streams}, but using the $\svtan$ values computed by including the systematic errors.}
\label{tab:table_systematic_r}
\begin{centering}
\begin{tabular}{|l|c|c|c|c|c|}
\hline
\hline
MW stream & in-situ & SCu & LCu & SCo & LCo\\
\hline
\hline
& & & & &\\

GD-1  &  0.289  ($ 0.773 $)  &  -4.37  ($ <10^{-3} $)  &  -10.05  ($ <10^{-3} $)  &  -1.314  ($ 0.189 $)  &  -1.218  ($ 0.223 $)\\

Fjorm  &  1.053  ($ 0.293 $)  &  -1.1  ($ 0.271 $)  &  -3.891  ($ <10^{-3} $)  &  0.198  ($ 0.843 $)  &  -0.122  ($ 0.903 $)\\

Phlegethon  &  0.791  ($ 0.429 $)  &  -2.068  ($ 0.039 $)  &  -5.701  ($ <10^{-3} $)  &  -0.378  ($ 0.705 $)  &  -0.617  ($ 0.537 $)\\

Sylgr  &  1.023  ($ 0.306 $)  &  -1.055  ($ 0.291 $)  &  -3.598  ($ <10^{-3} $)  &  0.124  ($ 0.901 $)  &  -0.164  ($ 0.869 $)\\

Gjoll  &  1.104  ($ 0.27 $)  &  -0.668  ($ 0.504 $)  &  -2.761  ($ 0.006 $)  &  0.379  ($ 0.704 $)  &  0.05  ($ 0.96 $)\\

& & & & & \\
\hline
\hline
\end{tabular}
\end{centering}

\end{table*}
\end{document}